\documentclass[aps, prd, amsfonts, amssymb, amsmath,
12pt, onecolumn, notitlepage, nofootinbib, showpacs]{revtex4-1}
\usepackage{graphicx}
\newcommand{\DP}{\Delta\Pi}
\newcommand{\ind}[2]{^{\text{\scriptsize $#1$}}_{\text{\scriptsize #2}}}
\newcommand{\inds}[2]{^{\text{\tiny $#1$}}_{\text{\tiny #2}}}
\newcommand{\RTau}[1]{R_{\tau, \text{\tiny #1}}}
\newcommand{\Nc}{N_{\text{\scriptsize c}}}
\newcommand{\nf}{n_{\text{\scriptsize f}}}
\newcommand{\Vud}{V_{\text{\scriptsize ud}}}
\newcommand{\Sew}{S_{\!\text{\tiny EW}}}
\newcommand{\DeltaQCD}[1]{\Delta^{\text{\tiny #1}}_{\text{\tiny QCD}}}
\newcommand{\dpew}{\delta'_{\text{\tiny EW}}}
\newcommand{\va}{_{\text{\tiny V/A}}}
\newcommand{\cva}{\chi\va}
\newcommand{\ML}{M_{l}}
\newcommand{\MTau}{M_{\tau}}

\newcommand{\includeplots}[2]{%
   \centerline{\includegraphics[width=80mm]{#1}%
   \hspace{6.4mm}%
   \includegraphics[width=80mm]{#2}}}

\begin{document}

\title{Dispersive approach to QCD and \\ inclusive $\tau$~lepton hadronic decay}

\author{A.V.~Nesterenko}
\email{nesterav@theor.jinr.ru}
\affiliation{Bogoliubov Laboratory of Theoretical Physics, \\
Joint Institute for Nuclear Research, \\
Dubna, 141980, Russian Federation}

\begin{abstract}
The dispersive approach to~QCD is applied to the study of the inclusive
$\tau$~lepton hadronic decay. This approach provides the unified integral
representations for the hadronic vacuum polarization function, related
$R$~function, and Adler function. These representations account for the
intrinsically nonperturbative constraints, which originate in the
kinematic restrictions on the functions on hand, and retain the effects
due to hadronization, which play a valuable role in the analysis of the
strong interaction processes at low energies. The dispersive approach
proves to be capable of describing recently updated ALEPH and OPAL
experimental data on inclusive $\tau$~lepton hadronic decay in vector and
axial--vector channels. The vicinity of values of the QCD scale parameter
obtained in both channels testifies to the potential ability of the
developed approach to describe the aforementioned data in a
self--consistent way.
\end{abstract}


\maketitle

\section{Introduction}
\label{Sect:Intro}

Hadronic vacuum polarization function~$\Pi(q^2)$ plays a central role in
various decisive tests of the self--consistency of Quantum
Chromodynamics~(QCD) and the entire Standard Model, that, in turn, puts
strong limits on possible new physics beyond the latter. In particular,
the theoretical description of such processes as inclusive $\tau$~lepton
hadronic decay, electron--positron annihilation into hadrons, as well as
the hadronic contributions to the muon anomalous magnetic moment and to
the running of the electromagnetic fine structure constant is inherently
based on~$\Pi(q^2)$. Additionally, the theoretical analysis of these
strong interaction processes constitutes a natural framework for a
thorough study of both perturbative and intrinsically nonperturbative
aspects of hadron dynamics.

The ultraviolet behavior of the hadronic vacuum polarization function can
reliably be calculated by making use of perturbation theory. However,
since the unambiguous method of the description of the strong interaction
processes at low energies is still far from being feasible, in order to
shed some light on~$\Pi(q^2)$ at low energies one inevitably resorts to a
variety of nonperturbative approaches. For instance, some hints on the
nonperturbative features of the hadronic vacuum polarization function can
be gained from such methods as lattice simulation~\cite{Lat1, Lat2, Lat3},
operator product expansion~\cite{OPE1, OPE2, OPE3, OPE4}, the instanton
liquid model~\cite{ILM1, ILM2}, and others.

One of the sources of the nonperturbative information about the
low--energy hadron dynamics is provided by dispersion relations.
Specifically, the latter renders the physical kinematic
restrictions\footnote{For example, the fact that the hadronic production
threshold has a nonvanishing value.} on the pertinent processes into the
mathematical form. As a result, dispersion relations impose stringent
constraints on relevant functions [such as~$\Pi(q^2)$, related~$R(s)$,
and~$D(Q^2)$; see Eqs.~(\ref{P_Def}), (\ref{R_Def}), and~(\ref{Adler_Def})
below], that should certainly be taken into account when one oversteps the
limits of perturbation theory. These constraints are embodied within the
\text{so--called} dispersive approach to QCD, which furnishes unified
integral representations for the functions on hand (see
Sec.~\ref{Sect:DQCD}).

The primary objective of this paper is to apply the dispersive approach to
QCD to the study of the inclusive $\tau$~lepton hadronic decay, that, in
particular, would allow one to properly account for the effects due to the
nonvanishing hadronic production threshold.

The layout of the paper is as follows. In Sec.~\ref{Sect:DQCD} the general
dispersion relations for the hadronic vacuum polarization function, the
so--called $R$~function, and Adler function are discussed
(Sec.~\ref{Sect:DR}) and the dispersive approach to~QCD is overviewed
(Sec.~\ref{Sect:DQCD_NR}). Section~\ref{Sect:Tau} deals with the inclusive
$\tau$~lepton hadronic decay. Specifically, Sec.~\ref{Sect:TauGen}
contains general remarks on this strong interaction process, whereas its
theoretical description is briefly recounted in Sec.~\ref{Sect:TauTheor}.
The analysis of recently updated ALEPH~\cite{ALEPH9805, ALEPH0608} and
OPAL~\cite{OPAL99, OPAL12} experimental data on inclusive $\tau$~lepton
hadronic decay within perturbative and dispersive approaches is performed
in Secs.~\ref{Sect:TauPert} and~\ref{Sect:TauDQCD}, respectively. In~the
Conclusions (Sec.~\ref{Sect:Concl}), the basic results are summarized and
further studies within this approach are outlined.

\section{Dispersive approach to Quantum Chromodynamics}
\label{Sect:DQCD}

\subsection{General dispersion relations for $\Pi(q^2)$, $R(s)$, and~$D(Q^2)$}
\label{Sect:DR}

\begin{figure*}[t]
\includeplots{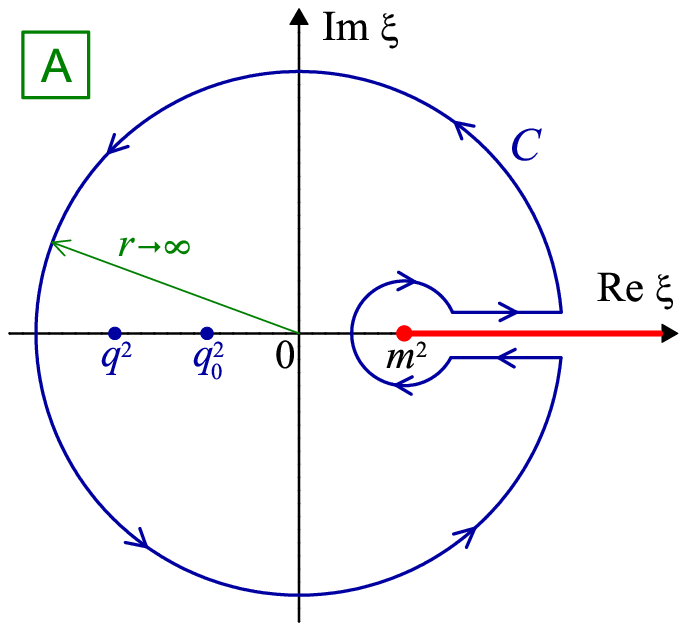}{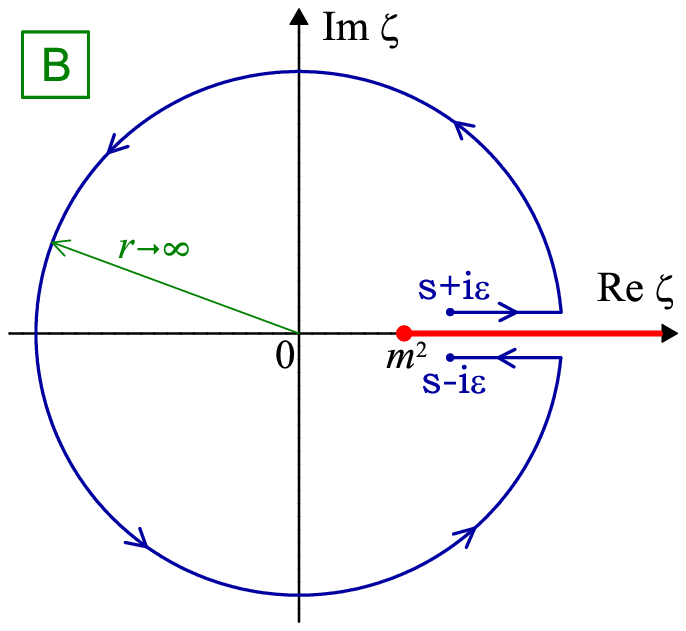}
\caption{A:~the closed integration contour~$C$ in Eq.~(\ref{P_Cauchy}).
The physical cut $\xi \ge m^2$ of the hadronic vacuum polarization
function $\Pi(\xi)$~(\ref{P_Def}) is shown along the positive semiaxis of
real~$\xi$. B:~the integration contour in Eq.~(\ref{R_Disp2}). The
physical cut $\zeta \ge m^2$ of the Adler function
$D(-\zeta)$~(\ref{Adler_Def}) is shown along the positive semiaxis of
real~$\zeta$.}
\label{Plot:PR_Contours}
\end{figure*}

As mentioned in the Introduction, the theoretical description of a number
of the strong interaction processes is inherently based on the hadronic
vacuum polarization function~$\Pi(q^2)$. The latter is defined as the
scalar part of the hadronic vacuum polarization tensor
\begin{eqnarray}
\label{P_Def}
\Pi_{\mu\nu}(q^2) &=& i\!\int\!d^4x\,e^{i q x} \langle 0 |\,
T\!\left\{J_{\mu}(x)\, J_{\nu}(0)\right\} | 0 \rangle
\nonumber \\
&=& \frac{i}{12\pi^2} (q_{\mu}q_{\nu} - g_{\mu\nu}q^2) \Pi(q^2).
\end{eqnarray}
For the processes involving final state hadrons the
function~$\Pi(q^2)$~(\ref{P_Def}) has the only cut along the positive
semiaxis of real~$q^2$ starting at the nonvanishing hadronic production
threshold~$q^2 \ge m^2$ (see discussion of this issue in, e.g.,
Ref.~\cite{Feynman}). In particular, the Feynman amplitude of the
respective process vanishes for the energies below the threshold, that
expresses the physical fact that the production of the final state hadrons
is kinematically forbidden for~$q^2<m^2$ (see also
Refs.~\cite{DQCDPrelim1, DQCD1a, DQCD1b} and references therein). Once the
location of the cut of function~$\Pi(q^2)$ in the complex $q^2$~plane is
known, one can write down the corresponding dispersion relation.
Specifically, bearing in mind the asymptotic ultraviolet behavior of the
hadronic vacuum polarization function, it is convenient to employ the
once--subtracted Cauchy integral formula:
\begin{equation}
\label{P_Cauchy}
\DP(q^2\!,\, q_0^2) = \frac{1}{2\pi i}\, (q^2-q_{0}^{2})
\oint_{\!C}^{}\! \frac{\Pi(\xi)}{(\xi - q^2)(\xi - q_{0}^{2})}\, d\xi.
\end{equation}
In this equation $\DP(q^2\!,\, q_0^2) = \Pi(q^2) - \Pi(q_0^2)$, whereas
the closed integration contour~$C$, which encloses both points~$q^2$
and~$q_0^2$ (commonly, one chooses~$q_0^2=0$) and goes counterclockwise
along the circle of the infinitely large radius, is displayed in
Fig.~\ref{Plot:PR_Contours}$\,$A. Equation~(\ref{P_Cauchy}) is usually
represented in the following form (see, e.g., Ref.~\cite{PDisp}):
\begin{equation}
\label{P_Disp}
\DP(q^2\!,\, q_0^2) = (q^2 - q_0^2) \int_{m^2}^{\infty}
\frac{R(\sigma)}{(\sigma-q^2)(\sigma-q_0^2)}\, d\sigma,
\end{equation}
where $R(s)$~stands for the discontinuity of the hadronic vacuum
polarization function across the physical cut
\begin{eqnarray}
\label{R_Def}
R(s) &=& \frac{1}{2 \pi i} \lim_{\varepsilon \to 0_{+}}
\Bigl[\Pi(s+i\varepsilon)-\Pi(s-i\varepsilon) \Bigr]
\nonumber \\
&=& \frac{1}{\pi}\, {\rm Im}\!\lim_{\varepsilon \to 0_{+}}\!
\Pi(s+i\varepsilon),
\end{eqnarray}
which is identified with the so--called $R$~ratio of electron--positron
annihilation into hadrons. It is worthwhile to note that in
Eq.~(\ref{R_Def}) the second equality holds for the functions~$\Pi(q^2)$
satisfying the condition~$\Pi(\xi^{*})=\Pi^{*}(\xi)$ only.

For practical purposes, it is convenient to define the Adler
function~\cite{Adler}
\begin{equation}
\label{Adler_Def}
D(Q^2) = - \frac{d\, \Pi(-Q^2)}{d \ln Q^2},
\end{equation}
with $Q^2 = -q^2 = -s > 0$ being the spacelike kinematic variable. It~is
worth mentioning that the subtraction point~$q_0^2$, appearing in
Eq.~(\ref{P_Disp}), does not enter in Eqs.~(\ref{R_Def})
and~(\ref{Adler_Def}). The dispersion relation for the Adler
function~$D(Q^2)$ follows directly from Eqs.~(\ref{P_Disp})
and~(\ref{Adler_Def}), namely~\cite{Adler}
\begin{equation}
\label{Adler_Disp}
D(Q^2) = Q^2 \int_{m^2}^{\infty} \frac{R(\sigma)}{(\sigma+Q^2)^2}\, d\sigma.
\end{equation}
In turn, the inverse relations, which express the functions~(\ref{P_Def})
and~(\ref{R_Def}) in terms of the Adler function, can be obtained by
integrating Eq.~(\ref{Adler_Def}) in finite limits, specifically
\begin{equation}
\label{R_Disp2}
R(s) =  \frac{1}{2 \pi i} \lim_{\varepsilon \to 0_{+}}
\int_{s + i \varepsilon}^{s - i \varepsilon}
D(-\zeta)\,\frac{d \zeta}{\zeta};
\end{equation}
see Refs.~\cite{R82, KP82}, and
\begin{equation}
\label{P_Disp2}
\DP(-Q^2\!,\, -Q_0^2) = - \int_{Q_0^2}^{Q^2} D(\zeta) \frac{d \zeta}{\zeta};
\end{equation}
see Ref.~\cite{C12}. The integration contour in Eq.~(\ref{R_Disp2}) lies
in the region of analyticity of the integrand; see
Fig.~\ref{Plot:PR_Contours}$\,$B.

It is worthwhile to outline that the derivation of
relations~(\ref{P_Disp}) and~(\ref{Adler_Disp})--(\ref{P_Disp2}) requires
only the knowledge of the location of the cut of hadronic vacuum
polarization function~$\Pi(q^2)$~(\ref{P_Def}) in the complex $q^2$~plane,
the asymptotic ultraviolet behavior of~$\Pi(q^2)$, and the
definitions~(\ref{R_Def}) and~(\ref{Adler_Def}). The derivation of
relations~(\ref{P_Disp}) and~(\ref{Adler_Disp})--(\ref{P_Disp2}) involves
neither additional approximations nor phenomenological assumptions.

As noted above, the aforementioned kinematic restrictions on the process
on hand are inherently embodied within corresponding dispersion relations.
In turn, the latter impose stringent physical intrinsically
nonperturbative constraints on the functions~$\Pi(q^2)$, $R(s)$,
and~$D(Q^2)$, that should certainly be accounted for when one comes out of
the limits of perturbation theory. For~example, Eq.~(\ref{P_Disp}) implies
that the hadronic vacuum polarization function~$\Pi(q^2)$ has the only cut
along the positive semiaxis of real~$q^2$ starting at the hadronic
production threshold~$q^2 \ge m^2$, whereas Eqs.~(\ref{R_Def})
and~(\ref{R_Disp2}) signify that the function~$R(s)$ acquires nonzero
values for real~$s$ above the threshold~($s \ge m^2$) only and accounts
for the effects\footnote{Such as the so--called ``$\pi^2$--terms''; see,
e.g., Refs.~\cite{pi2terms1a, pi2terms1b, pi2terms2}, Sec.~2.5 of
review~\cite{Prosperi}, as well as Ref.~\cite{pi2terms3}.} due to
continuation of spacelike theoretical results into the timelike domain.
In~turn, Eq.~(\ref{Adler_Disp}) implies that the Adler function
vanishes\footnote{This condition holds for~$m^2 \neq 0$ only.} in the
infrared limit [\text{$D(Q^2) \to 0$} at \text{$Q^2 \to 0$}] and possesses
the only cut along the negative semiaxis of real~$Q^2$ starting at the
hadronic production threshold~\text{$Q^2 \le -m^2$} (see
Refs.~\cite{DQCD1a, DQCD2, C12} and references therein).

\subsection{Novel integral representations for
$\Pi(q^2)$, $R(s)$, and~$D(Q^2)$}
\label{Sect:DQCD_NR}

Equations~(\ref{P_Disp})--(\ref{P_Disp2}) constitute the complete set of
relations, which express the functions~$\Pi(q^2)$, $R(s)$, and~$D(Q^2)$ in
terms of each other. For practical purposes, it proves to be convenient to
deal with the unified integral representations, which express the
functions on hand in terms of the common spectral density~$\rho(\sigma)$.
Such representations are obtained in the framework of the so--called
dispersive approach to QCD (see Refs.~\cite{DQCDPrelim1, DQCD1a, DQCD2,
C12} and references therein).

In particular, the integral representation for the
function~$R(s)$~(\ref{R_Def}) can be derived from Eq.~(\ref{R_Disp2}) (the
proper integration contour is displayed in
Fig.~\ref{Plot:PR_Contours}$\,$B) and the fact that the strong correction
to the Adler function vanishes in the ultraviolet asymptotic:
\begin{equation}
\label{R_DQCD}
R(s) = R^{(0)}(s) + \theta(s-m^2) \int_{s}^{\infty}\!
\rho(\sigma) \frac{d\,\sigma}{\sigma}.
\end{equation}
In this equation~$R^{(0)}(s)$ denotes the leading--order (i.e.,~zeroth
order in the strong running coupling) term of the function~$R(s)$,
$\theta(x)$ is the unit step~function [$\theta(x)=1$ if $x \ge 0$ and
$\theta(x)=0$ otherwise], and~$\rho(\sigma)$ stands for the spectral
density specified in Eq.~(\ref{RhoGen}) below. In turn, integral
representation for the hadronic vacuum polarization function~(\ref{P_Def})
can be obtained\footnote{Derivation of Eq.~(\ref{P_DQCD}) from
Eqs.~(\ref{P_Disp}) and~(\ref{R_DQCD}) involves the integration by parts.}
by making use of Eqs.~(\ref{P_Disp}) and~(\ref{R_DQCD})
\begin{eqnarray}
\label{P_DQCD}
\DP(q^2,\, q_0^2) &=& \DP^{(0)}(q^2,\, q_0^2)
\nonumber \\
&& +\!\int_{m^2}^{\infty} \rho(\sigma)
\ln\biggl(\frac{\sigma-q^2}{\sigma-q_0^2}
\frac{m^2-q_0^2}{m^2-q^2}\biggr)\frac{d\,\sigma}{\sigma}, \qquad
\end{eqnarray}
whereas integral representation for the Adler function~(\ref{Adler_Def})
can be derived directly from Eqs.~(\ref{Adler_Def}) and~(\ref{P_DQCD})
\begin{equation}
\label{Adler_DQCD}
D(Q^2) = D^{(0)}(Q^2) + \frac{Q^2}{Q^2+m^2}
\int_{m^2}^{\infty} \rho(\sigma)
\frac{\sigma-m^2}{\sigma+Q^2} \frac{d\,\sigma}{\sigma}. \quad
\end{equation}
The spectral density appearing in Eqs.~(\ref{R_DQCD})--(\ref{Adler_DQCD})
reads
\begin{eqnarray}
\label{RhoGen}
\rho(\sigma) &=& \frac{1}{2 \pi i} \frac{d}{d\,\ln\sigma}
\lim_{\varepsilon \to 0_{+}}
\Bigl[p(\sigma-i\varepsilon)-p(\sigma+i\varepsilon) \Bigr] \qquad
\nonumber \\
&=& - \frac{d}{d\,\ln\sigma}\, r(\sigma)
\nonumber \\
&=& \frac{1}{2 \pi i} \lim_{\varepsilon \to 0_{+}}
\Bigl[d(-\sigma-i\varepsilon)-d(-\sigma+i\varepsilon) \Bigr]\!.
\end{eqnarray}
Here~$p(q^2)$, $r(s)$, and~$d(Q^2)$ denote the strong corrections to the
functions~$\Pi(q^2)$, $R(s)$, and~$D(Q^2)$, respectively. For the
functions~$p(q^2)$ and~$d(Q^2)$ satisfying conditions \text{$p(\xi^{*}) =
p^{*}(\xi)$} and~$d(\xi^{*}) = d^{*}(\xi)$ Eq.~(\ref{RhoGen}) acquires the
form
\begin{eqnarray}
\label{RhoGen2}
\rho(\sigma) &=& \frac{1}{\pi} \frac{d}{d\,\ln\sigma}\,
\text{Im}\lim_{\varepsilon \to 0_{+}} p(\sigma-i\varepsilon)
\nonumber \\
&=& - \frac{d}{d\,\ln\sigma}\, r(\sigma)
\nonumber \\
&=& \frac{1}{\pi}\, \text{Im}\lim_{\varepsilon \to 0_{+}}
d(-\sigma-i\varepsilon).
\end{eqnarray}
It is straightforward to verify that the
functions~\text{(\ref{R_DQCD})--(\ref{Adler_DQCD})} satisfy all the
relations~\text{(\ref{P_Disp})--(\ref{P_Disp2})}. In~particular, the
latter means that the representation~(\ref{R_DQCD}) can also be obtained
from Eqs.~(\ref{R_Def}) and~(\ref{P_DQCD}), the
representation~(\ref{Adler_DQCD}) can also be derived from
Eqs.~(\ref{Adler_Disp}) and~(\ref{R_DQCD}),~etc. The discussion of this
issue can be found in Refs.~\cite{DQCD1a, DQCD1b, DQCD2, DQCD3, DQCD4,
C12} and references therein.

The leading--order terms in Eqs.~(\ref{R_DQCD})--(\ref{Adler_DQCD}) have
the following form~\cite{Feynman, QEDAB}:
\begin{subequations}
\begin{eqnarray}
\DP^{(0)}(q^2,\, q_0^2) &=& \frac{2}{\tan^3\varphi}(\varphi - \tan\varphi)
\nonumber \\
&& - \frac{2}{\tan^3\varphi_{0}}(\varphi_{0} - \tan\varphi_{0}), \\
R^{(0)}(s) &=& \theta(s - m^2)\biggl(1-\frac{m^2}{s}\biggr)^{\!\!3/2}, \\
D^{(0)}(Q^2) &=& 1 + \frac{3}{\xi}\Bigl[1 \!-\! \sqrt{1\!+\!\xi^{-1}}\,
\sinh^{-1}\!\bigl(\xi^{1/2}\bigr)\!\Bigr]\!,~ \qquad
\end{eqnarray}
\end{subequations}
where $\sin^2\varphi = q^2/m^2$, $\sin^2\varphi_{0} = q^{2}_{0}/m^2$, and
$\xi=Q^2/m^2$; see also Refs.~\cite{DQCD3, DQCD4, C12}. It~is worth
mentioning here that a rough approximation for the leading--order terms of
the functions~(\text{\ref{R_DQCD})--(\ref{Adler_DQCD}}) (the so--called
``abrupt kinematic threshold''),
\begin{subequations}
\begin{eqnarray}
\label{PRD_AKT}
\DP\inds{(0)}{AKT}(q^2,\, q_0^2) &=& -\ln\biggl(\frac{m^2-q^2}{m^2-q_0^2}\biggr), \\
R\inds{(0)}{AKT}(s) &=& \theta(s-m^2), \\
D\inds{(0)}{AKT}(Q^2) &=& \frac{Q^2}{Q^2+m^2},
\end{eqnarray}
\end{subequations}
which, nonetheless, grasps the basic peculiarities of the functions on
hand, was discussed in Refs.~\cite{DQCD1a, DQCD1b, DQCD2, DQCD3, C12}.

\begin{figure}[t]
\includegraphics[width=80mm]{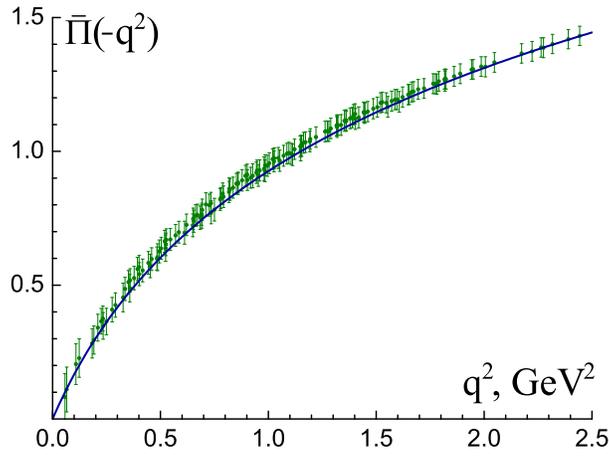}
\caption{Comparison of the hadronic vacuum polarization
function~(\ref{P_DQCD}) [$\bar{\Pi}(q^2)=\DP(0,q^2)$, solid curve] with
relevant lattice simulation data~\cite{Lat2} (circles). The presented
results correspond to the spectral density~(\ref{RhoDef}) and $\nf=2$
active flavors; see Refs.~\cite{NPQCD13, Prep}.}
\label{Plot:PDQCD}
\end{figure}

The integral representations~(\ref{R_DQCD})--(\ref{Adler_DQCD})
automatically embody all the nonperturbative
constraints\footnote{Including the correct analytic properties in the
kinematic variable, which implies that the
functions~\text{(\ref{R_DQCD})--(\ref{Adler_DQCD})} are free of unphysical
singularities.}, which Eqs.~(\ref{P_Disp})--(\ref{P_Disp2}) impose on the
functions on hand (see Sec.~\ref{Sect:DR}). It is worthwhile to note that
a preliminary formulation of the dispersive approach to QCD, which
accounts for only one of the aforementioned constraints on the Adler
function (namely, the cut~\text{$Q^2 \le -m^2$} along the negative
semiaxis of real~$Q^2$), was discussed in Refs.~\cite{DQCDPrelim1,
DQCDPrelim2}. The integral
representations~(\ref{R_DQCD})--(\ref{Adler_DQCD}) were obtained by
employing only the relations~\text{(\ref{P_Disp})--(\ref{P_Disp2})} and
the asymptotic ultraviolet behavior of the hadronic vacuum polarization
function. Neither additional approximations nor phenomenological
assumptions were involved in the derivation of
Eqs.~(\ref{R_DQCD})--(\ref{Adler_DQCD}). As one can infer from
Fig.~\ref{Plot:PDQCD}, the hadronic vacuum polarization
function~(\ref{P_DQCD}) is in a good agreement with relevant low--energy
lattice simulation data~\cite{Lat2}; see Refs.~\cite{NPQCD13, Prep}. It~is
worth mentioning also that the Adler function~(\ref{Adler_DQCD}) complies
with the corresponding experimental prediction in the entire energy range
(see, in particular, Refs.~\cite{DQCD1a, DQCD1b, DQCD2}), and the
representations~(\ref{R_DQCD})--(\ref{Adler_DQCD}) conform with the
results obtained in Ref.~\cite{PRL99PRD77}.

So far, there is no method to restore the unique complete expression for
the spectral density~$\rho(\sigma)$~(\ref{RhoGen}) (discussion of this
issue may be found in, e.g., Refs.~\cite{DQCD2, PRD62PRD64, Review, C12}).
Nonetheless, the perturbative contribution to~$\rho(\sigma)$ can be
calculated by making use of the perturbative expression for either of the
strong corrections appearing in Eq.~(\ref{RhoGen}) (see, e.g.,
Ref.~\cite{CPC}):
\begin{eqnarray}
\label{RhoPert}
\rho\ind{}{pert}(\sigma) &=& \frac{1}{2 \pi i} \frac{d}{d\,\ln\sigma}
\lim_{\varepsilon \to 0_{+}}
\Bigl[p\ind{}{pert}(\sigma-i\varepsilon) -
p\ind{}{pert}(\sigma+i\varepsilon) \Bigr]
\nonumber \\
&=&  - \frac{d}{d\,\ln\sigma}\, r\ind{}{pert}(\sigma)
\nonumber \\
&=& \frac{1}{2 \pi i} \lim_{\varepsilon \to 0_{+}}
\Bigl[d\ind{}{pert}(-\sigma-i\varepsilon) -
d\ind{}{pert}(-\sigma+i\varepsilon) \Bigr]\!.
\end{eqnarray}
In this paper the following model for the spectral density will
be employed (see also Refs.~\cite{DQCD3, DQCD4, C12}):
\begin{equation}
\label{RhoDef}
\rho(\sigma) = \frac{4}{\beta_{0}}\frac{1}{\ln^{2}(\sigma/\Lambda^2)+\pi^2} +
\frac{\Lambda^2}{\sigma}.
\end{equation}
Expression~(\ref{RhoDef}) represents a simplest ansatz for~$\rho(\sigma)$,
which merges the one--loop perturbative contribution [first term on the
right--hand side of Eq.~(\ref{RhoDef})] with an intrinsically
nonperturbative (i.e.,~containing inverse power of~$\sigma$) term, and
involves a minimal number of parameters. In this way (likewise to other
similar models~\cite{APT, PRD62PRD64}), the scale parameter~$\Lambda$
remains the only adjustable quantity. It should be noted that the
contribution of the last term of Eq.~(\ref{RhoDef}) to the asymptotic
ultraviolet behavior of the Adler function~(\ref{Adler_DQCD}) is of the
form~$\exp(-1/a)$ with~$a=\alpha\ind{(1)}{pert}\beta_{0}/(4\pi)$ being the
one--loop perturbative ``couplant''. This fact implies that the second
term on the right--hand side of Eq.~(\ref{RhoDef}) gives no contribution
to the expansion of the Adler function~(\ref{Adler_DQCD}) in powers of~$a$
at~$a \to 0_{+}$ (i.e.,~at~$Q^2\to\infty$) and does not alter its
perturbative approximation~(\ref{AdlerPert}).

Note that in the massless limit~($m^2=0$) the integral
representations~(\ref{R_DQCD})--(\ref{Adler_DQCD}) acquire
the form
\begin{subequations}
\begin{eqnarray}
\label{P_DQCD0}
\DP(q^2,\, q_0^2) &=& \DP\ind{(0)}{pert}(q^2,\, q_0^2)
\nonumber \\
&& + \!\int_{0}^{\infty}\! \rho(\sigma)
\ln\!\Biggl[\frac{1-(\sigma/q^2)}{1-(\sigma/q_0^2)}\Biggr]
\frac{d\,\sigma}{\sigma}, \qquad \\
\label{R_DQCD0}
R(s) &=& \theta(s) \biggl[R\ind{(0)}{pert}(s) +  \int_{s}^{\infty}
\!\rho(\sigma) \frac{d\,\sigma}{\sigma} \biggr]\!, \\
\label{Adler_DQCD0}
D(Q^2) &=& D\ind{(0)}{pert}(Q^2) + \int_{0}^{\infty}
\frac{\rho(\sigma)}{\sigma+Q^2}\, d\,\sigma,
\end{eqnarray}
\end{subequations}
where the leading--order terms read
\begin{subequations}
\label{PRD_Pert0}
\begin{eqnarray}
\DP\ind{(0)}{pert}(q^2,\, q_0^2) &=& -\ln\Biggl(\frac{-q^2}{-q_0^2}\Biggr)\!, \\
R\ind{(0)}{pert}(s) &=& 1, \\
D\ind{(0)}{pert}(Q^2) &=& 1.
\end{eqnarray}
\end{subequations}
It is worthwhile to mention that for the case of perturbative spectral
density [$\rho(\sigma) = \text{Im}\; d\ind{}{pert}(-\sigma -
i\,0_{+})/\pi$] two massless equations~(\ref{R_DQCD0})
and~(\ref{Adler_DQCD0}) become identical to those of the so--called
analytic perturbation theory (APT)~\cite{APT} (see also Refs.~\cite{APT1,
APT2, APT3, APT4, APT5, APT6}).

However, it is essential to keep the value of the hadronic production
threshold nonvanishing. Specifically, whereas in the ultraviolet
asymptotic the effects due to hadronization (i.e.,~due to \text{$m^2 \neq
0$}) can be safely neglected, in the infrared domain such effects become
substantial and play a valuable role in the studies of the strong
interaction processes at low energies. In particular, as it has been noted
in Sec.~\ref{Sect:DR}, the massless limit $(m^2 = 0)$ loses some of the
intrinsically nonperturbative constraints, which relevant dispersion
relations impose on the functions on hand. For example, the difference
between the representation~(\ref{Adler_DQCD}) and its massless
limit~(\ref{Adler_DQCD0}) was elucidated in Sec.~4 of Ref.~\cite{DQCD1a}
and Sec.~3 of Ref.~\cite{DQCD1b}.

\section{Inclusive $\tau$~lepton hadronic decay}
\label{Sect:Tau}

\subsection{General remarks}
\label{Sect:TauGen}

The inclusive $\tau$~lepton hadronic decay is characterized by the
experimentally measurable ratio of two widths:
\begin{equation}
\label{RTauExpDef}
R_{\tau} = \frac{\Gamma(\tau^{-} \to \text{hadrons}^{-}\,
\nu_{\tau})} {\Gamma(\tau^{-} \to e^{-}\, \bar{\nu}_{e}\, \nu_{\tau})}.
\end{equation}
This inclusive semileptonic branching ratio is usually decomposed into
several parts, specifically
\begin{equation}
\label{RTauExp}
R_{\tau} =
\RTau{V}^{\text{\tiny $J$=0}} + \RTau{V}^{\text{\tiny $J$=1}} +
\RTau{A}^{\text{\tiny $J$=0}} + \RTau{A}^{\text{\tiny $J$=1}} + \RTau{S}.
\end{equation}
On the right--hand side of this equation the first four terms account for
the hadronic decay modes involving light quarks~($u$,~$d$) only and
associated with vector~(V) and axial--vector~(A) quark currents,
respectively. The last term on the right--hand side of Eq.~(\ref{RTauExp})
accounts for the $\tau$~lepton hadronic decay modes that involve the
strange quark. The superscript~$J$ in Eq.~(\ref{RTauExp}) indicates the
angular momentum in the hadronic rest frame.

\begin{figure*}[t]
\includeplots{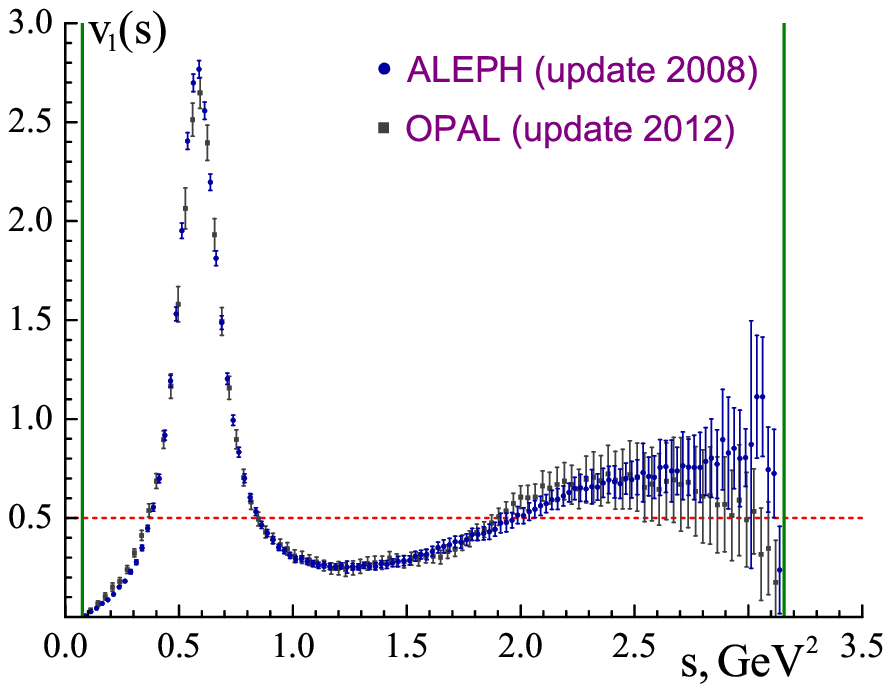}{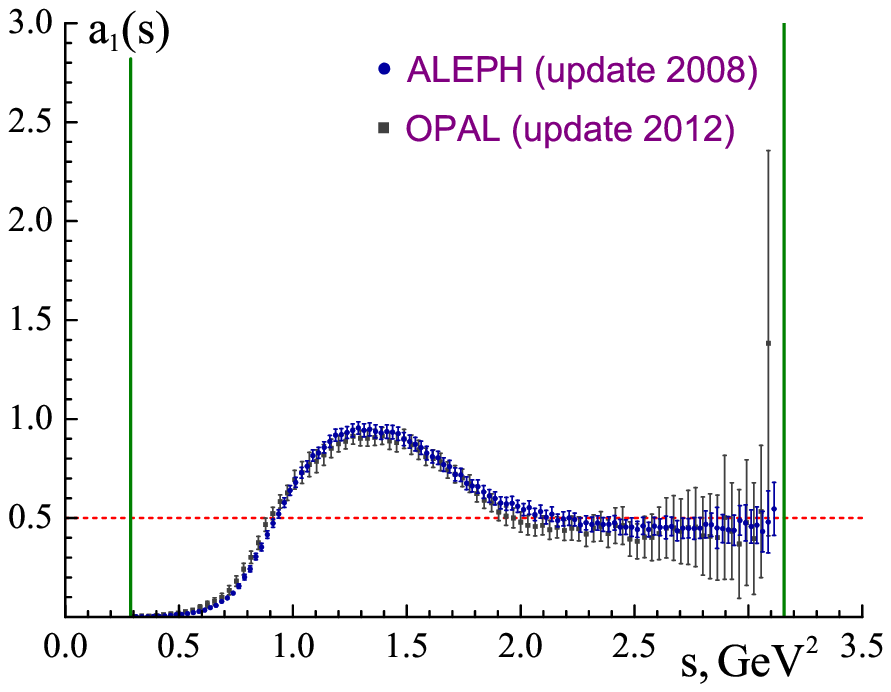}
\caption{The inclusive $\tau$~lepton hadronic decay vector (left--hand
plot) and axial--vector (right--hand plot) spectral functions. The
experimental data~\cite{ALEPH0608} (update of the ALEPH
measurement~\cite{ALEPH9805}) and~\cite{OPAL12} (update of the OPAL
measurement~\cite{OPAL99}) are shown by circles and boxes, respectively.
Vertical solid lines mark the boundaries of respective kinematic
intervals, whereas horizontal dashed lines denote the naive massless
parton model prediction.}
\label{Plot:TauSpFun}
\end{figure*}

Basically, the evaluation of the quantities appearing in
Eq.~(\ref{RTauExp}) involves the so--called spectral functions, which are
extracted from the experiment. For the zero angular momentum ($J=0$) the
vector spectral function vanishes (that yields~$\RTau{V}^{\text{\tiny
$J$=0}}=0$), whereas the axial--vector one is commonly approximated by the
Dirac $\delta$~function, since the main contribution is due to the pion
pole here. The experimental predictions for the nonstrange spectral
functions corresponding to $J=1$ by ALEPH~\cite{ALEPH9805, ALEPH0608} and
OPAL~\cite{OPAL99, OPAL12} collaborations are displayed in
Fig.~\ref{Plot:TauSpFun}. In what follows we shall restrict ourselves to
the study of the terms~$\RTau{V}^{\text{\tiny $J$=1}}$
and~$\RTau{A}^{\text{\tiny $J$=1}}$ of $R_{\tau}$~ratio~(\ref{RTauExp}).

The theoretical expression for the aforementioned quantities reads
\begin{equation}
\label{RTauTheor}
\RTau{V/A}^{\text{\tiny $J$=1}} = \frac{\Nc}{2}\,|\Vud|^2\,\Sew\,
\Bigl(\DeltaQCD{V/A} + \dpew \Bigr),
\end{equation}
where $\Nc=3$ is the number of colors, $|\Vud| = 0.97425 \pm 0.00022$ is
the Cabibbo--Kobayashi--Maskawa matrix element~\cite{PDG2012}, $\Sew =
1.0194 \pm 0.0050$ and $\dpew = 0.0010$ stand for the electroweak
corrections~\cite{EWF}, and
\begin{equation}
\label{DeltaQCDDef}
\DeltaQCD{V/A} = 2\int_{m\va^2}^{\ML^2}\!
\biggl(1-\frac{s}{\ML^2}\biggr)^{\!\!2}\biggl(1+2\frac{s}{\ML^2}\biggr) R(s)\,
\frac{d s}{\ML^2}
\end{equation}
denotes the hadronic contribution; see Ref.~\cite{BNP}. The
function~$R(s)$ appearing in the integrand of Eq.~(\ref{DeltaQCDDef}) is
defined in Eq.~(\ref{R_Def}). The experimental predictions for the
functions~$\DeltaQCD{V/A}$~(\ref{DeltaQCDDef}) corresponding to the
recently updated ALEPH~\cite{ALEPH0608} and OPAL~\cite{OPAL12} data are,
respectively,
\begin{subequations}
\label{DeltaQCDExp}
\begin{eqnarray}
\label{DeltaQCDExpA08}
\Delta\ind{\text{\tiny V}}{exp} = 1.224 \pm 0.050, &\qquad&
\Delta\ind{\text{\tiny A}}{exp} = 0.748 \pm 0.034, \\
\label{DeltaQCDExpO12}
\Delta\ind{\text{\tiny V}}{exp} = 1.229 \pm 0.088, &\qquad&
\Delta\ind{\text{\tiny A}}{exp} = 0.741 \pm 0.058.
\end{eqnarray}
\end{subequations}

It is worthwhile to note that in Eq.~(\ref{DeltaQCDDef}) $\ML$~denotes the
mass of the lepton on hand, whereas $m\va$~stands for the value of the
hadronic production threshold (i.e., the total mass of the lightest
allowed hadronic decay mode of this~lepton in the corresponding channel).
The nonvanishing value of~$m\va$, which exceeds the masses of two lightest
leptons, explicitly expresses the physical fact that the $\tau$~lepton is
the only lepton that is heavy enough (\mbox{$\MTau \simeq
1.777\,$GeV$\,$\cite{PDG2012}}) to decay into hadrons. Specifically, in
the massless limit ($m\va=0$) the theoretical
prediction~(\ref{DeltaQCDPert}) for the hadronic
contribution~(\ref{DeltaQCDDef}) to Eq.~(\ref{RTauTheor}) is nonvanishing
for either lepton (\mbox{$l = e, \mu, \tau$}). In particular, the
leading--order term of Eq.~(\ref{DeltaQCDPert})
(\mbox{$\Delta\ind{(0)}{pert}=1$}), which corresponds to the naive
massless parton model prediction~(\ref{PRD_Pert0}), does not depend
on~$\ML$, and, therefore, is the same for any lepton. In~the realistic
case (i.e., when the total mass of the lightest allowed hadronic decay
mode exceeds the masses of electron and muon, $M_{e}<M_{\mu}<m\va<\MTau$)
Eq.~(\ref{DeltaQCDDef}) acquires nonzero value for the case of the
$\tau$~lepton only (discussion of this issue can also be found in
Refs.~\cite{DQCD4, C12} and references therein).

\subsection{Theoretical evaluation of~$\Delta^{\text{\normalfont\tiny V/A}}_{\text{\normalfont\tiny QCD}}$}
\label{Sect:TauTheor}

\begin{figure*}[t]
\includeplots{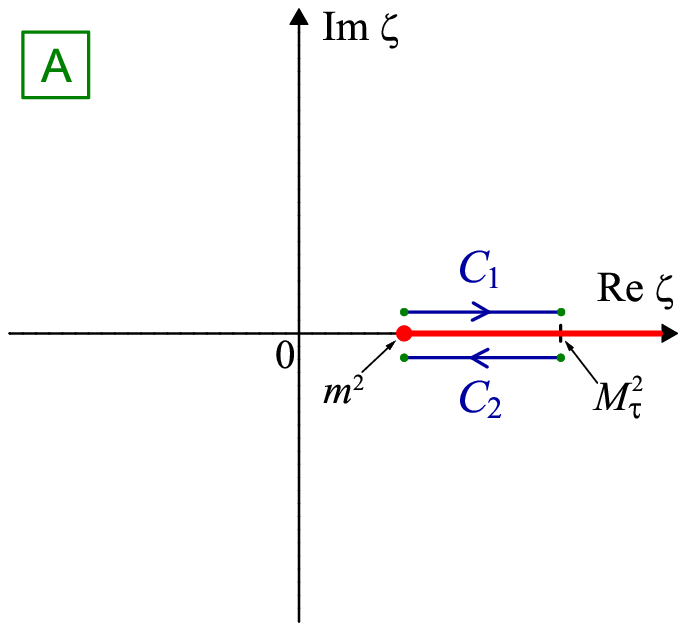}{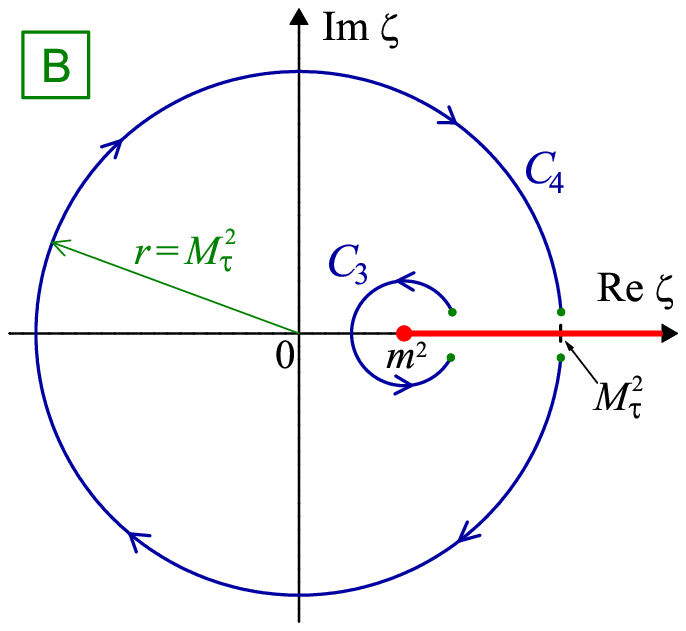}
\caption{The integration contour~$C_{1}+C_{2}$ in
Eq.~(\ref{DeltaQCD1})~(A) and its continuous
deformation~$C_{3}+C_{4}$~(B). The physical cut $\zeta \ge m^2$ of the
Adler function $D(-\zeta)$~(\ref{Adler_Def}) is shown along the positive
semiaxis of real~$\zeta$.}
\label{Plot:RTau_Contours}
\end{figure*}

Theoretical analysis of the hadronic contribution~(\ref{DeltaQCDDef}) to
Eq.~(\ref{RTauTheor}) usually begins with the integration by parts, which
casts Eq.~(\ref{DeltaQCDDef}) into the form\footnote{Derivation of
Eq.~(\ref{DeltaQCD1}) from Eq.~(\ref{DeltaQCDDef}) employs
Eq.~(\ref{R_Disp2}).} (the indices ``V'' and~``A'' will only be shown when
relevant hereinafter)
\begin{eqnarray}
\label{DeltaQCD1}
\DeltaQCD{} &=& g(1) R(\MTau^2) - g(\chi) R(m^2)
\nonumber \\
&& + \frac{1}{2\pi i} \int\limits_{C_{1}+C_{2}} g\biggl(\frac{\zeta}{\MTau^2}\biggr)
D(-\zeta) \frac{d \zeta}{\zeta}.
\end{eqnarray}
Here the functions~$R(s)$ and~$D(Q^2)$ are defined in Eqs.~(\ref{R_Def})
and~(\ref{Adler_Def}), respectively, $\chi = m^2/\MTau^2$, and
\begin{equation}
\label{g_Def}
g(x) = x (2 - 2x^2 + x^3).
\end{equation}
The piecewise continuous integration contour appearing in the last term of
Eq.~(\ref{DeltaQCD1}) is displayed in Fig.~\ref{Plot:RTau_Contours}$\,$A.
Specifically, the integration contour~$C_{1}+C_{2}$ consists of two
straight lines, which go from~$m^2+i\varepsilon$ to~$\MTau^2+i\varepsilon$
and from~$\MTau^2-i\varepsilon$ to~$m^2-i\varepsilon$ (the
limit~$\varepsilon \to 0_{+}$ is assumed in what follows). If the Adler
function~$D(Q^2)$ appearing in the integrand of the last term of
Eq.~(\ref{DeltaQCD1}) possesses the correct analytic properties in the
kinematic variable~$Q^2$ (see Sec.~\ref{Sect:DR}), then the integration
contour~$C_{1}+C_{2}$ can be continuously deformed into the integration
contour~$C_{3}+C_{4}$ shown in Fig.~\ref{Plot:RTau_Contours}$\,$B:
\begin{eqnarray}
\label{DeltaQCD1a}
\DeltaQCD{} &=& g(1) R(\MTau^2) - g(\chi) R(m^2)
\nonumber \\
&& + \frac{1}{2\pi i} \int\limits_{C_{3}+C_{4}} g\biggl(\frac{\zeta}{\MTau^2}\biggr)
D(-\zeta) \frac{d \zeta}{\zeta}.
\end{eqnarray}
Here the integration contour~$C_{3}$ is the nonclosed circle of vanishing
radius, which goes counterclockwise from~$m^2+i\varepsilon$
to~$m^2-i\varepsilon$, whereas the integration contour~$C_{4}$ is the
nonclosed circle of radius~$\MTau^2$, which goes clockwise
from~$\MTau^2-i\varepsilon$ to~$\MTau^2+i\varepsilon$.

Despite the remarks given in Sec.~\ref{Sect:DQCD}, the massless limit
($m=0$) will be adopted in the rest of this subsection and in
Sec.~\ref{Sect:TauPert}. Since the function $g(x)$~(\ref{g_Def}) vanishes
at~$x \to 0$, the second term in Eq.~(\ref{DeltaQCD1a}) and the integral
along the contour~$C_{3}$ (which is centered at~$\zeta=0$ in the massless
limit) in the last term of Eq.~(\ref{DeltaQCD1a}) do not
contribute\footnote{The regular behavior of functions~$R(s)$ and~$D(Q^2)$
at the threshold is assumed here.} to~$\DeltaQCD{}$, which takes the
following form in this case:
\begin{equation}
\label{DeltaQCD2}
\DeltaQCD{} = R(\MTau^2) +
\frac{1}{2\pi i} \int\limits_{C_{4}}
g\biggl(\frac{\zeta}{\MTau^2}\biggr)
D(-\zeta) \frac{d \zeta}{\zeta}.
\end{equation}
The first term of this equation can be represented in the form of
Eq.~(\ref{R_Disp2}) with the integration contour~$C_4$ shown in
Fig.~\ref{Plot:RTau_Contours}$\,$B, that (after appropriate change
of the integration variable) leads to
\begin{equation}
\label{DeltaQCD4}
\DeltaQCD{} = \frac{1}{2\pi} \lim_{\varepsilon \to 0_{+}}
\int\limits_{-\pi+\varepsilon}^{\pi-\varepsilon}
\Bigl[1 - g\bigl(-e^{i\theta}\bigr)\!\Bigr]
D\Bigl(\MTau^2e^{i\theta}\Bigr) d \theta.
\end{equation}

\subsection{Inclusive $\tau$~lepton hadronic decay within perturbative approach}
\label{Sect:TauPert}

From the very beginning, it is necessary to outline that what was obtained
in the previous subsection, Eq.~(\ref{DeltaQCD4}), is only valid for the
massless limit of the Adler function~$D(Q^2)$, which possesses the
correct\footnote{Otherwise, Eq.~(\ref{DeltaQCD4}) cannot be derived from
Eq.~(\ref{DeltaQCDDef}).} analytic properties in the kinematic
variable~$Q^2$. However, in the framework of the perturbative approach one
commonly directly employs in Eq.~(\ref{DeltaQCD4}) the perturbative
approximation~$D\ind{}{pert}(Q^2)$, which has unphysical singularities
in~$Q^2$. Specifically, at the $\ell$--loop level
\begin{equation}
\label{AdlerPert}
D\ind{(\ell)}{pert}(Q^2) = D\ind{(0)}{pert}(Q^2) + \sum_{j=1}^{\ell} d_{j}
\Bigl[\alpha\ind{(\ell)}{pert}(Q^2)\Bigr]^{j}.
\end{equation}
In this equation at the one--loop level (i.e., for \mbox{$\ell=1$}) the
strong running coupling reads $\alpha\ind{(1)}{pert}(Q^2) =
4\pi/[\beta_{0}\,\ln(Q^2/\Lambda^2)]$, where $\beta_{0}=11-2\nf/3$,
$\Lambda$~denotes the QCD scale parameter, $\nf$~stands for the number of
active flavors, and~$d_{1}=1/\pi$; see Refs.~\cite{AdlerPert4La,
AdlerPert4Lb} and references therein. In what follows the one--loop level
with $\nf=3$ active flavors will be assumed.

\begin{figure*}[t!]
\includeplots{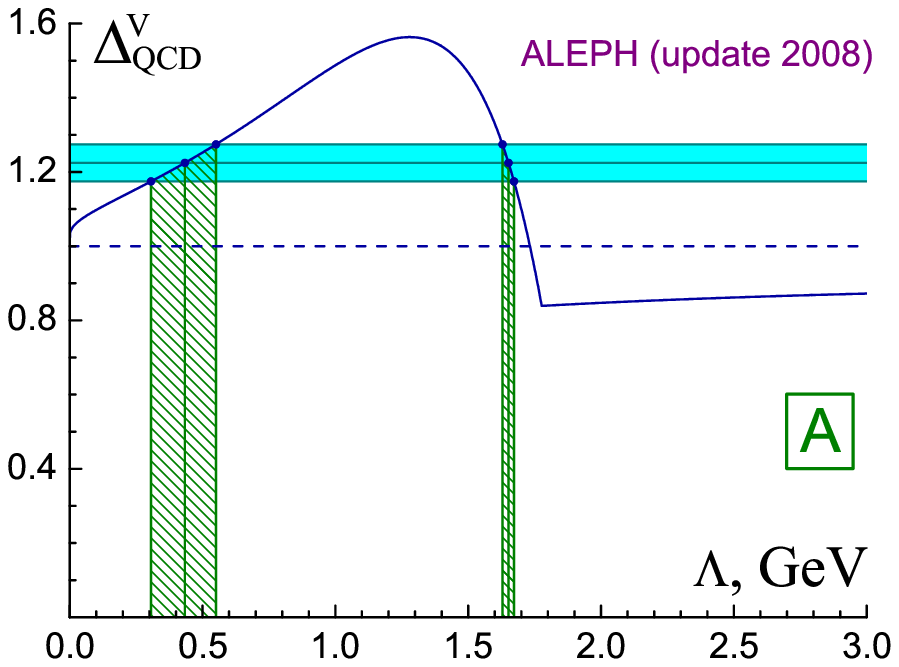}{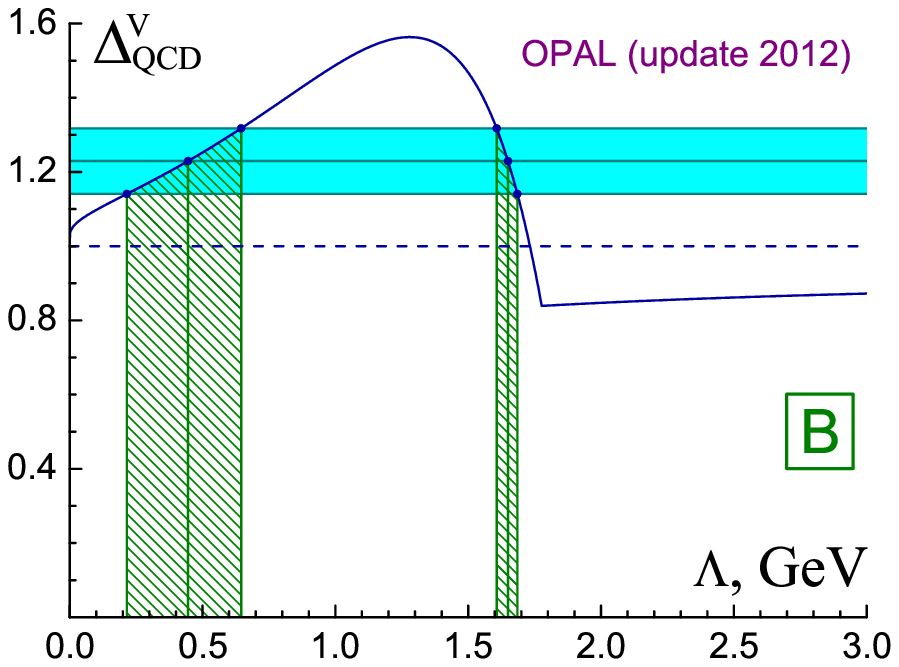}
\vskip5.5mm
\includeplots{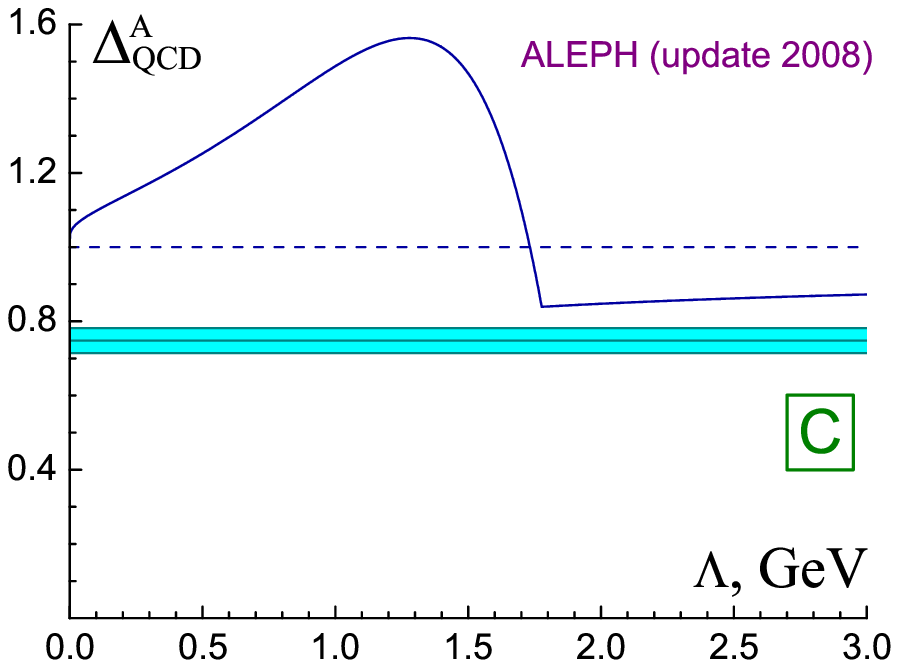}{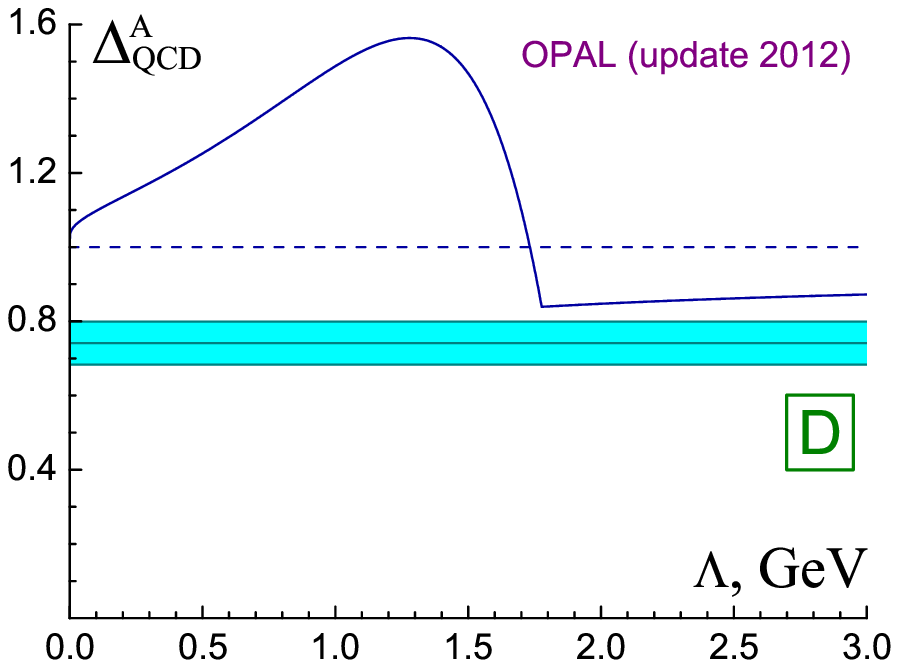}
\caption{Comparison of the perturbative expression~$\Delta\ind{\text{\tiny
V/A}}{pert}$~(\ref{DeltaQCDPert}) (solid curves) with relevant
experimental data. The plots A,~C and B,~D correspond to
Eq.~(\ref{DeltaQCDExpA08}) (Ref.~\cite{ALEPH0608}) and
Eq.~(\ref{DeltaQCDExpO12}) (Ref.~\cite{OPAL12}), respectively. The
leading--order term~$\Delta\ind{(0)}{pert}$~(\ref{DeltaQCDPert}) is shown
by the horizontal dashed line, whereas the solution for the QCD scale
parameter~$\Lambda$ (if it exists) is denoted by the vertical dashed
band.}
\label{Plot:RTauPT}
\end{figure*}

Thus, the substitution of the one--loop perturbative expression for the
Adler function~(\ref{AdlerPert}) to Eq.~(\ref{DeltaQCD4}) eventually leads
to
\begin{equation}
\label{DeltaQCDPert}
\Delta\ind{\text{\tiny V/A}}{pert} = \Delta\ind{(0)}{pert} +
\frac{4}{\beta_{0}}\!\int_{0}^{\pi}
\frac{\lambda A_{1}(\theta)+\theta A_{2}(\theta)}{\pi(\lambda^2+\theta^2)}
\,d\theta.
\end{equation}
In this equation~$\Delta\ind{(0)}{pert}=1$,
$\lambda = \ln \bigl( \MTau^2/\Lambda^2 \bigr)$, and
\begin{eqnarray}
A_{1}(\theta) &=& 1 + 2\cos(\theta) - 2\cos(3\theta) - \cos(4\theta), \\
A_{2}(\theta) &=& 2\sin(\theta) - 2\sin(3\theta) - \sin(4\theta).
\end{eqnarray}
Note that the leading--order term on the right--hand side of
Eq.~(\ref{DeltaQCDPert}) (\mbox{$\Delta\ind{(0)}{pert}=1$}) corresponds to
a rather rough massless perturbative approximation of the functions on
hand~(\ref{PRD_Pert0}), which is applicable, in fact, in the ultraviolet
asymptotic only. Besides, this leading--order term~$\Delta\ind{(0)}{pert}$
appears to be independent of any of the involved kinematic parameters (see
also discussion of this issue in Sec.~\ref{Sect:TauGen} and
Refs.~\cite{DQCD4, C12}).

\begin{table*}[t]
\caption{Values of the QCD scale parameter~$\Lambda\,$[MeV] obtained
within perturbative approach [see Eq.~(\ref{DeltaQCDPert})
and~Fig.~\ref{Plot:RTauPT}].}
\label{Tab:RTauPT}
\begin{ruledtabular}
\begin{tabular}{lcc}
\hspace{37.5mm}
& ~ALEPH data [Eq.~(\ref{DeltaQCDExpA08}), Ref.~\cite{ALEPH0608}]~~
& ~~OPAL data [Eq.~(\ref{DeltaQCDExpO12}), Ref.~\cite{OPAL12}]~ \\
\hline
Vector channel
& \underline{{\scriptsize\strut}$434_{-127}^{+117}$} \qquad $1652_{-23}^{+21}$
& \underline{{\scriptsize\strut}$445_{-230}^{+201}$} \qquad $1650_{-43}^{+36}$ \\
Axial--vector channel & $\cdots$ & $\cdots$ \\
\end{tabular}
\end{ruledtabular}
\end{table*}

It is worth emphasizing that the perturbative approach provides identical
expressions~(\ref{DeltaQCDPert}) for the functions~(\ref{DeltaQCDDef}) in
vector and axial--vector channels (i.e., \mbox{$\Delta\ind{\text{\tiny
V}}{pert} \equiv \Delta\ind{\text{\tiny A}}{pert}$}). However, their
experimental values~\cite{ALEPH9805, ALEPH0608, OPAL99, OPAL12} specified
in Eq.~(\ref{DeltaQCDExp}) are different (i.e.,
\mbox{$\Delta\ind{\text{\tiny V}}{exp} \neq \Delta\ind{\text{\tiny
A}}{exp}$}). The~juxtaposition of the perturbative
expression~(\ref{DeltaQCDPert}) with its experimental
predictions~(\ref{DeltaQCDExp}) is presented in Fig.~\ref{Plot:RTauPT} and
the obtained results are listed in Table~\ref{Tab:RTauPT}. As one can
infer from Figs.~\ref{Plot:RTauPT}$\,$A and~\ref{Plot:RTauPT}$\,$B, for
the vector channel there are two\footnote{Usually, for the vector channel
the underlined value of~$\Lambda$ given in Table~\ref{Tab:RTauPT} is
retained, whereas the other value is considered as a formal solution and
merely disregarded.} solutions for the QCD scale parameter~$\Lambda$. As
for the axial--vector channel, the experimental data~\cite{ALEPH9805,
ALEPH0608, OPAL99, OPAL12} cannot be described within the perturbative
approach. In particular, the use of the massless limit results in the fact
that the leading--order term of Eq.~(\ref{DeltaQCDPert}) far exceeds the
corresponding experimental prediction~$\Delta\ind{\text{\tiny A}}{exp}$.
Specifically, as one can infer from Figs.~\ref{Plot:RTauPT}$\,$C
and~\ref{Plot:RTauPT}$\,$D, for any value of the QCD scale
parameter~$\Lambda$ the function~$\Delta\ind{\text{\tiny
A}}{pert}$~(\ref{DeltaQCDPert}) lies above~$\Delta\ind{\text{\tiny
A}}{exp}$ specified in Eq.~(\ref{DeltaQCDExp}).

\subsection{Inclusive $\tau$~lepton hadronic decay within dispersive approach}
\label{Sect:TauDQCD}

It is crucial to emphasize that the effects due to the nonvanishing value
of the hadronic production threshold have been completely left out in the
massless limit\footnote{It is worthwhile to mention that there is a number
of papers, which study the inclusive semileptonic branching
ratio~(\ref{RTauExp}) within massless APT and its modifications; see
Refs.~\cite{TauAPT1, TauAPT2, TauAPT3}. However, these papers basically
deal either with the total sum of vector and axial--vector terms of
Eq.~(\ref{RTauExp}) or with the vector term of Eq.~(\ref{RTauExp}) only.}
examined above. However, as it was outlined in Sec.~\ref{Sect:DQCD}, such
effects play a significant role in the studies of the strong interaction
processes at low energies. The dispersive approach to QCD described in
Sec.~\ref{Sect:DQCD} properly accounts for the effects due to
hadronization and embodies the aforementioned intrinsically
nonperturbative constraints on the functions on hand. In the analysis of
the inclusive $\tau$~lepton hadronic decay presented in this
subsection\footnote{Description of the inclusive $\tau$~lepton hadronic
decay given in Ref.~\cite{DQCDPrelim1} corresponds to the preliminary
formulation of the dispersive approach to QCD, which does not account for
some of the nonperturbative constraints on the functions on hand; see
Sec.~\ref{Sect:DQCD}.} the integral representations obtained within the
dispersive approach to~QCD~(\ref{R_DQCD})--(\ref{Adler_DQCD}) will be
employed and the hadronic production threshold will be kept nonvanishing.

\begin{figure*}[t]
\includeplots{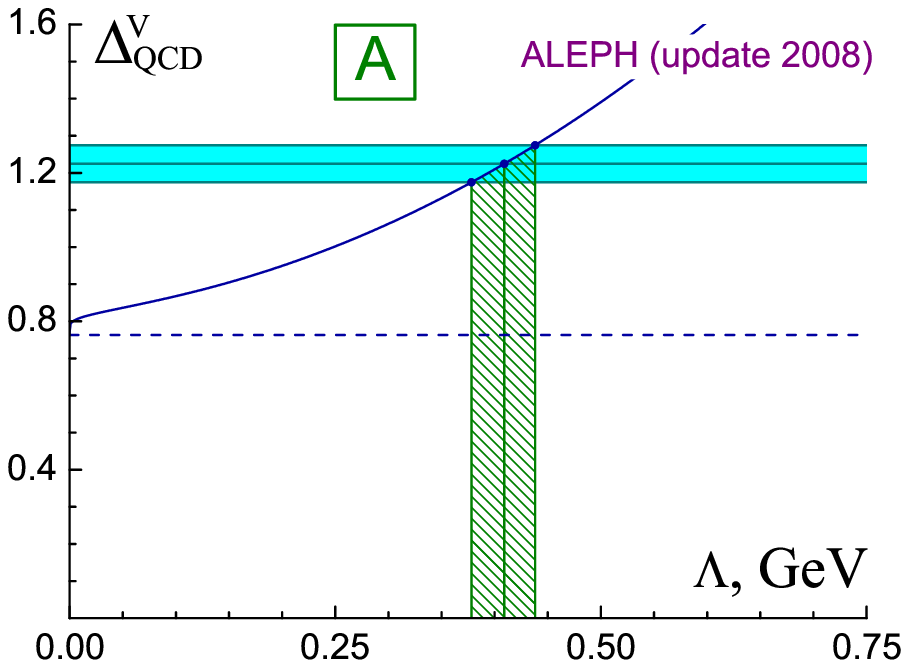}{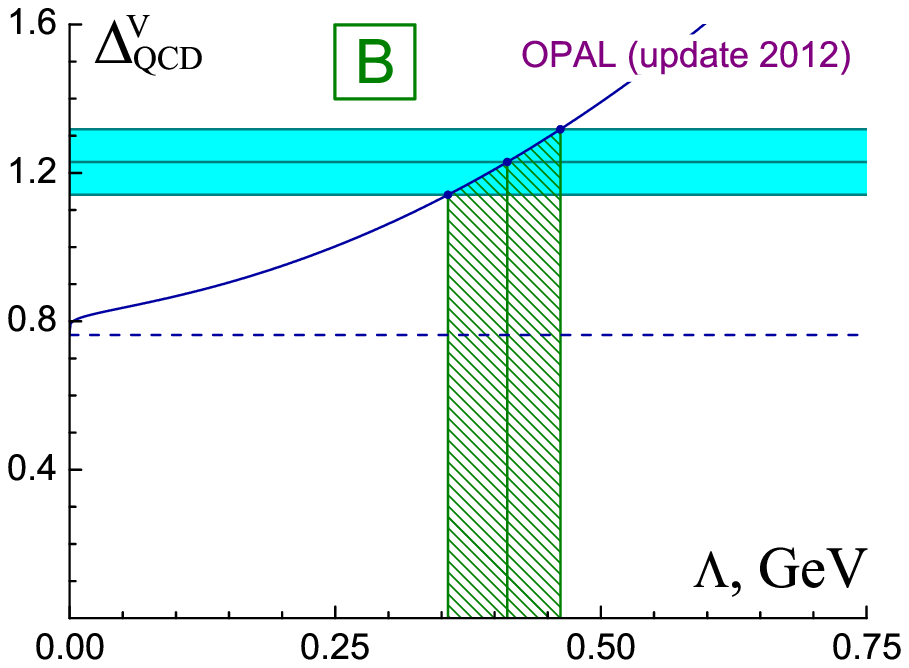}
\vskip5.5mm
\includeplots{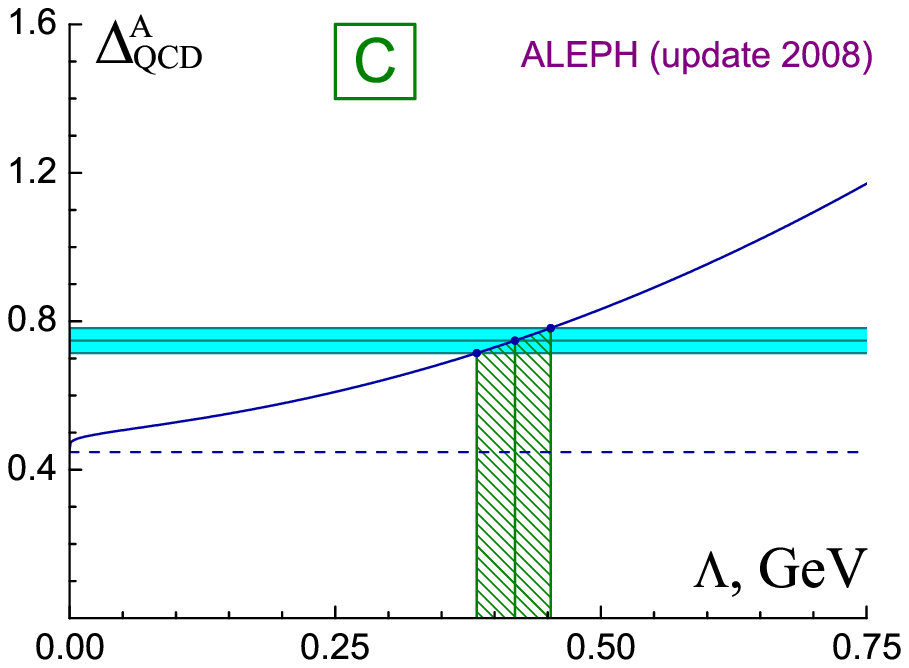}{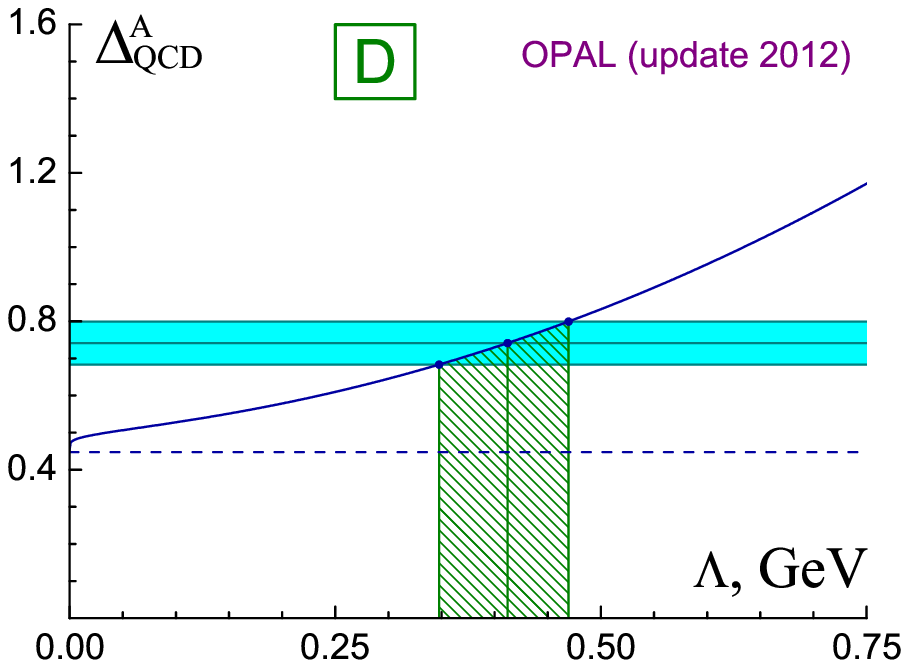}
\caption{Comparison of the
expression~$\DeltaQCD{V/A}$~(\ref{DeltaQCD_DQCD}) (solid curves) with
relevant experimental data. The plots A,~C and B,~D correspond to
Eq.~(\ref{DeltaQCDExpA08}) (Ref.~\cite{ALEPH0608}) and
Eq.~(\ref{DeltaQCDExpO12}) (Ref.~\cite{OPAL12}), respectively. The
leading--order term~(\ref{DeltaQCD_DQCD}) is shown by the horizontal
dashed line, whereas the solution for the QCD scale parameter~$\Lambda$ is
denoted by the vertical dashed band.}
\label{Plot:RTauDQCD}
\end{figure*}

As noted above, the functions~(\ref{R_DQCD})--(\ref{Adler_DQCD}) possess
the correct analytic properties in the kinematic variable. Hence, their
use with any equivalent transformation of the initial
expression~(\ref{DeltaQCDDef}) of the hadronic
contribution~$\DeltaQCD{V/A}$ to Eq.~(\ref{RTauTheor}) leads to the same
result. For~instance, one can use the representation~(\ref{R_DQCD}) in
Eq.~(\ref{DeltaQCDDef}), as well as the representations~(\ref{R_DQCD})
and~(\ref{Adler_DQCD}) in Eq.~(\ref{DeltaQCD1a}). Eventually, in the
framework of the dispersive approach to QCD the hadronic
contribution~(\ref{DeltaQCDDef}) to Eq.~(\ref{RTauTheor}) acquires the
following form:
\begin{eqnarray}
\label{DeltaQCD_DQCD}
\DeltaQCD{V/A} &=& 3\,g_{1}\Bigl(\frac{\cva}{2}\Bigr)\sqrt{1-\cva}
\nonumber \\
&& - 3\,g_{2}\Bigl(\frac{\cva}{4}\Bigr)
\ln\biggl(\sqrt{\cva^{-1}}+\sqrt{\cva^{-1}-1}\biggr) \quad
\nonumber \\
&& + \int_{m\va^{2}}^{\infty}\!G\Bigl(\frac{\sigma}{M_{\tau}^{2}}\Bigr)\,
\rho(\sigma)\,\frac{d \sigma}{\sigma}\,,
\end{eqnarray}
where $\chi\va = m\va^{2}/\MTau^{2}$,
$m_{\text{\tiny V}}^{2} \simeq 0.075\,\text{GeV}^2$,
$m_{\text{\tiny A}}^{2} \simeq 0.288\,\text{GeV}^2$,
$G(x) = g(x)\,\theta(1-x) + g(1)\,\theta(x-1) - g(\chi\va)$,
spectral density~$\rho(\sigma)$ is specified in Eq.~(\ref{RhoDef}),
function~$g(x)$ is defined in Eq.~(\ref{g_Def}), and
\begin{eqnarray}
g_{1}(x) &=& \frac{1}{3} + 4x -\frac{5}{6}x^2 + \frac{1}{2}x^3, \\
g_{2}(x) &=& 8x(1 + 2x^2 - 2x^3);
\end{eqnarray}
see also Refs.~\cite{DQCD3, DQCD4, C12} and references therein.

\begin{table*}[t]
\caption{Values of the QCD scale parameter~$\Lambda\,$[MeV] obtained
within dispersive approach [see Eq.~(\ref{DeltaQCD_DQCD})
and~Fig.~\ref{Plot:RTauDQCD}].}
\label{Tab:RTauDQCD}
\begin{ruledtabular}
\begin{tabular}{lcc}
\hspace{37.5mm}
& ~ALEPH data [Eq.~(\ref{DeltaQCDExpA08}), Ref.~\cite{ALEPH0608}]~~
& ~~OPAL data [Eq.~(\ref{DeltaQCDExpO12}), Ref.~\cite{OPAL12}]~ \\
\hline
Vector channel & $408 \pm 30$ & $409 \pm 53$ \\
Axial--vector channel & $418 \pm 35$ & $409 \pm 61$ \\
\end{tabular}
\end{ruledtabular}
\end{table*}

The juxtaposition of the obtained result~(\ref{DeltaQCD_DQCD}) with
pertinent experimental predictions~(\ref{DeltaQCDExp}) is presented in
Fig.~\ref{Plot:RTauDQCD} and the corresponding values of the QCD scale
parameter~$\Lambda$ are given in Table~\ref{Tab:RTauDQCD}. As one can
infer from Fig.~\ref{Plot:RTauDQCD}, the dispersive approach is capable of
describing the experimental data on inclusive $\tau$~lepton hadronic
decay~\cite{ALEPH9805, ALEPH0608, OPAL99, OPAL12} in vector and
axial--vector channels. The respective values of~$\Lambda$ conform to the
one reported in the previous subsection. Additionally, the values of the
QCD scale parameter obtained in vector and axial--vector channels appear
to be nearly identical to each other.

It is worth noting also that there is still no method to restore the
unique complete expression for the spectral density~$\rho(\sigma)$ (see
Sec.~\ref{Sect:DQCD}). This fact implies that, in general, within the
approach on hand (identically to other similar approaches\footnote{In
particular, as mentioned in Sec.~V of Ref.~\cite{DQCDPrelim1}, the
analysis of the vector channel data on inclusive $\tau$~lepton hadronic
decay within massless~APT yields a considerably larger value of~$\Lambda$
for the spectral density~\cite{APT} (see also Ref.~\cite{TauAPT1}) than
for the spectral density~\cite{PRD62PRD64}.}) the ensuing value
of~$\Lambda$ depends on the particular choice of~$\rho(\sigma)$.
Specifically, the values of the scale parameter listed in
Table~\ref{Tab:RTauDQCD} correspond to the model~(\ref{RhoDef}).
Nonetheless, the vicinity of values of~$\Lambda$ obtained in vector and
axial--vector channels testifies to the potential ability of the developed
approach to describe the experimental data~\cite{ALEPH9805, ALEPH0608,
OPAL99, OPAL12} in a self--consistent way.

\section{Conclusions}
\label{Sect:Concl}

The dispersive approach to QCD is applied to the study of the inclusive
$\tau$~lepton hadronic decay. This approach enables one to retain the
effects due to hadronization, which appear to be valuable in the
low--energy domain, and to account for the intrinsically nonperturbative
constraints, which originate in the kinematic restrictions on the process
on hand. The obtained results indicate that the dispersive approach is
capable of describing recently updated ALEPH~\cite{ALEPH9805, ALEPH0608}
and OPAL~\cite{OPAL99, OPAL12} experimental data on inclusive
$\tau$~lepton hadronic decay in vector and axial--vector channels. The
values of QCD scale parameter~$\Lambda$ evaluated in both channels are
nearly identical to each other, which bears witness to the potential
ability of the developed approach to describe the experimental
data~\cite{ALEPH9805, ALEPH0608, OPAL99, OPAL12} in a self--consistent
way.

In further studies it would certainly be interesting to include into the
presented analysis of inclusive $\tau$~lepton hadronic decay the higher
order perturbative corrections and nonperturbative contributions arising
from the operator product expansion, as well as to explore possible
constraints on the spectral density appearing in the approach on hand.

\begin{acknowledgments}
The author is grateful to D.~Boito, P.~Colangelo, M.~Davier, F.~De~Fazio,
S.~Menke, and H.~Wittig for stimulating discussions and useful comments.
The partial support of Grant No.~JINR-12-301-01 is also acknowledged.
\end{acknowledgments}

\end{document}